%% file: ms.tex
\documentclass[preprint]{aastex6}

\usepackage{amsmath}

\begin{document}

\title{Stellar Parameters for Pulsating B Star Candidates in the Kepler Field}

\author{Richard J.\ Hanes\altaffilmark{1}, Steven Waskie\altaffilmark{1}, Jonathan M.\ Labadie-Bartz\altaffilmark{1,2}, Audrey Wall\altaffilmark{1,3}, Amber Boyer\altaffilmark{4}, M.\ Virginia McSwain\altaffilmark{1}}
\altaffiltext{1}{Department of Physics, Lehigh University, 16 Memorial Drive East, Bethlehem, PA 18015, USA}
\altaffiltext{2}{Current Address: Department of Physics \& Astronomy, University of Delaware, Newark, DE 19716, USA}
\altaffiltext{3}{Current Address: 56 Sparta Ave, Newton, NJ 07860, USA}
\altaffiltext{4}{Department of Physical Sciences, Kutztown University of Pennsylvania, 15200 Kutztown Rd, Kutztown PA 19530}

\begin{abstract}

The field of asteroseismology has enjoyed a large swath of data coming from recent missions (e.g., \textit{CoRoT}, \textit{Kepler}, \textit{K2}). This wealth of new data has allowed the field to expand beyond the previous limitation of a few extremely bright and evolved stars. Asteroseismology relies on accurate surface measurements for boundary conditions, but the predicted physical parameters in the Kepler Input Catalog (KIC) are unreliable for hot stars.  We present stellar parameters of 25 candidate pulsating B star candidates in the Kepler field. We use blue optical spectra to measure the projected rotational velocity ($V \sin i$), effective temperature ($T_{\rm eff}$), and surface gravity ($\log g$) using TLUSTY and Kurucz ATLAS9 model atmospheres.  We find a large discrepancy between our spectroscopically derived parameters and those derived from photometry in the KIC and \textit{Gaia} Data Release 2 (DR2). Using spectral energy distributions, we also measure the radii of these stars and later calculate the luminosities and masses. We find the extinctions ($A_V$) of these stars to be consistent with zero, which is expected for stars of high Galactic latitude.
\end{abstract}

\keywords{stars: massive -- stars: oscillations -- stars: fundamental parameters}

\section{Introduction} \label{sec:intro}

Although helioseismology has long proven to be an invaluable tool for calibrating models of the solar interior, general asteroseismological analysis was previously limited to a few extremely bright or evolved stars \citep{chaplin2013}. With the wide-field, high-precision, \textit{Kepler} and \textit{K2} missions, astronomers are now able to perform asteroseismology on tens of thousands of stars \citep{chaplin2011,Yu2018} across a broad range of temperatures and evolutionary status. Even in its final months, \textit{K2} has continued to provide insight into the pulsations of its target stars. 

Despite the high Galactic latitude of the original \textit{Kepler} field, 85 pulsating B-type candidates have been identified in that field \citep{Balona2011, mcnamara2012}.  They exhibit a range of nonradial pulsations (NRPs) with periods consistent with $\beta$ Cephei variables, slowly pulsating B stars (SPB), and hybrids between those two classes.  The precision light curves available with \textit{Kepler} are proving to reveal many high frequency and low-amplitude modes, and with excellent frequency resolution, that are not detectable from the ground.  

Asteroseismology of B stars with NRPs is currently being used to improve stellar structure and evolutionary models for hot stars (e.g., \citealt{saesen2010}). Their various pulsation frequencies probe different layers of their interiors. Doing so, however requires accurate boundary conditions at the stellar surface (e.g., \citealt{huber2012}). Spectroscopy allows accurate measurements of effective temperature, $T_{\rm eff}$, and surface gravity, $\log g$, that are essential in constraining stellar radii, ages, and evolutionary spin-down rates. Knowledge of the projected rotational velocity, $V \sin i$, is key for studying the angular momentum of NRPs.

One hurdle for the asteroseismic analysis of B-type pulsators in the \textit{Kepler} field is the lack of accurate physical parameters for these stars. The KIC uses the SDSS $g - r$ color as a temperature indicator, but the Rayleigh-Jeans slope of the hot star spectral energy distributions means that the $g - r$ color is largely insensitive to temperature for B-type stars.  The KIC photometric $\log g$ measurements are likewise poor since the $g-r$ index does not sample the Balmer jump, which is strongly dependent on atmospheric pressure and thus $\log g$. \cite{Balona2011} used spectroscopic line profile fitting to measure $T_{\rm eff}$ of 30 B stars in the \textit{Kepler} field and found substantial differences from the predicted $T_{\rm eff}$ of the same stars in the KIC (\citealt{brown2011}).  \cite{pinsonneault2012} published revised temperature scales for the KIC, but only for stars with $4,000 {\; \rm K} \le T_{\rm eff} \le 7,000$ K, which is substantially cooler than the stars considered in this work. 

We present here the results of the measurements of $T_{\rm eff}$, $\log g$, and $V \sin i$ of 25 candidate $\beta$ Cephei, SPB, and hybrid pulsating B stars in the \textit{Kepler} field with $8 \le V \le 16$. Section 2 details our observations and data reduction of the spectra. In Section 3, we describe our measurements of $T_{\rm eff}$, $\log g$, and $V \sin i$ of these stars using the Tlusty BSTAR2006 grid and Kurucz ATLAS9 model atmospheres. Comparing $T_{\rm eff}$ and $\log g$ to the evolutionary tracks of \cite{ekstrom2012}, we also measure the mass, radius, and age. Section 4 compares our results with the KIC and the \textit{Gaia} Data Release 2 as well as other published works. Calculated distances from \cite{Jones2018} using the Gaia parallaxes are also included in order to estimate extinctions ($A_V$) for these stars.

\section{Observations} \label{sec:observations}

We observed each target using the KPNO 4m Mayall telescope with the RC spectrograph from 2014 May 9-13.  We used the grating BL 380 in $2^{nd}$ order, a $\rm CaSO_4$ order sorting filter, a 1.5 arcsec slit, and the T2KA CCD to achieve resolving power $R = \lambda / \Delta \lambda \approx 7,200$.  With a central wavelength of 4340 \AA, this setup allowed us to observe the range 4,060--4,620 \AA, covering several useful helium and hydrogen lines.  We reduced the raw spectra with the \textsc{doslit} package of IRAF. All spectra were wavelength calibrated using an FeAr arc lamp.

\section{Spectral Modeling} \label{sec:modeling}

Two grids of synthetic spectra were used in our modeling process to measure $T_{\rm eff}$, $\log g$, and $V \sin i$. First, we used a grid of line blanketed, plane-parallel, local thermodynamic equilibrium (LTE) models generated using the ATLAS9 code \citep{kurucz1994} for stars with $T_{\rm eff} < 15,000$ K. The non-LTE (NLTE) Tlusty BSTAR2006 \citep{lanz2007} model spectra were used for stars with $T_{\rm eff} > 15,000$ K. We adopted grids of (Z/Z$_\odot$ = 1) and a microturbulent velocity of V$_t$ = 2 km s$^{-1}$.

Before fitting, we estimated $T_{\rm eff}$ and $\log g$ of the stars based on the strength and shapes of the Balmer and helium lines. We measured the projected rotational velocity ($V \sin i$) by using custom IDL codes to compare the observed profiles of \ion{He}{2} $\lambda$4026, \ion{He}{1} $\lambda\lambda$4387, 4471,  and \ion{Mg}{2} $\lambda$4481 to limb darkened, rotationally broadened, and instrumentally broadened model profiles using steps of 10 km s$^{-1}$. For each step, we compared $\Sigma$(O-C)$^2$,the sum of the squares of the residuals, and determined the minimal value of a parabolic fit as the value for $V \sin i$. The error in $V \sin i$ was determined by allowing a 5\% tolerance in $\Sigma$(O-C)$^2$. Table \ref{tab:vsini} lists the measurements of $V \sin i$ for all of the helium lines as well as their weighted averages.

We then modeled H$\gamma$ lines for $T_{\rm eff}$ and $\log g$ using broadened models according to our measured $V \sin i$ along each point in our generated ATLAS9 grid or BSTAR2006 gird. Once we found the closest match with the grid, we used linear interpolation between the grid points to find the best fit for $T_{\rm eff}$ and $\log g$. The errors in $T_{\rm eff}$ and $\log g$ were determined by allowing 5\% tolerance of the $\Sigma$(O-C)$^2$. Our $T_{\rm eff}$ and $\log g$ measurements are recorded in Table \ref{tab:mm}. 

\section{Discussion} \label{sec:discussion}

As expected, we find large discrepancies between photometrically derived $T_{\rm eff}$ and $\log g$ and our measurements. In columns 2 and 3 of Table \ref{tab:Param_comp}, we show the derived $T_{\rm eff}$ and $\log g$  from the KIC for our observed stars \citep{brown2011}. We include in column 4 of Table \ref{tab:Param_comp} the $T_{\rm eff}$ from the recently released DR2 \citep{Andrae2018}. Columns 7 and 8 of this table give our measurements using spectroscopic fitting as well as the uncertainties for these values.  
 
\begin{figure}
\gridline{\fig{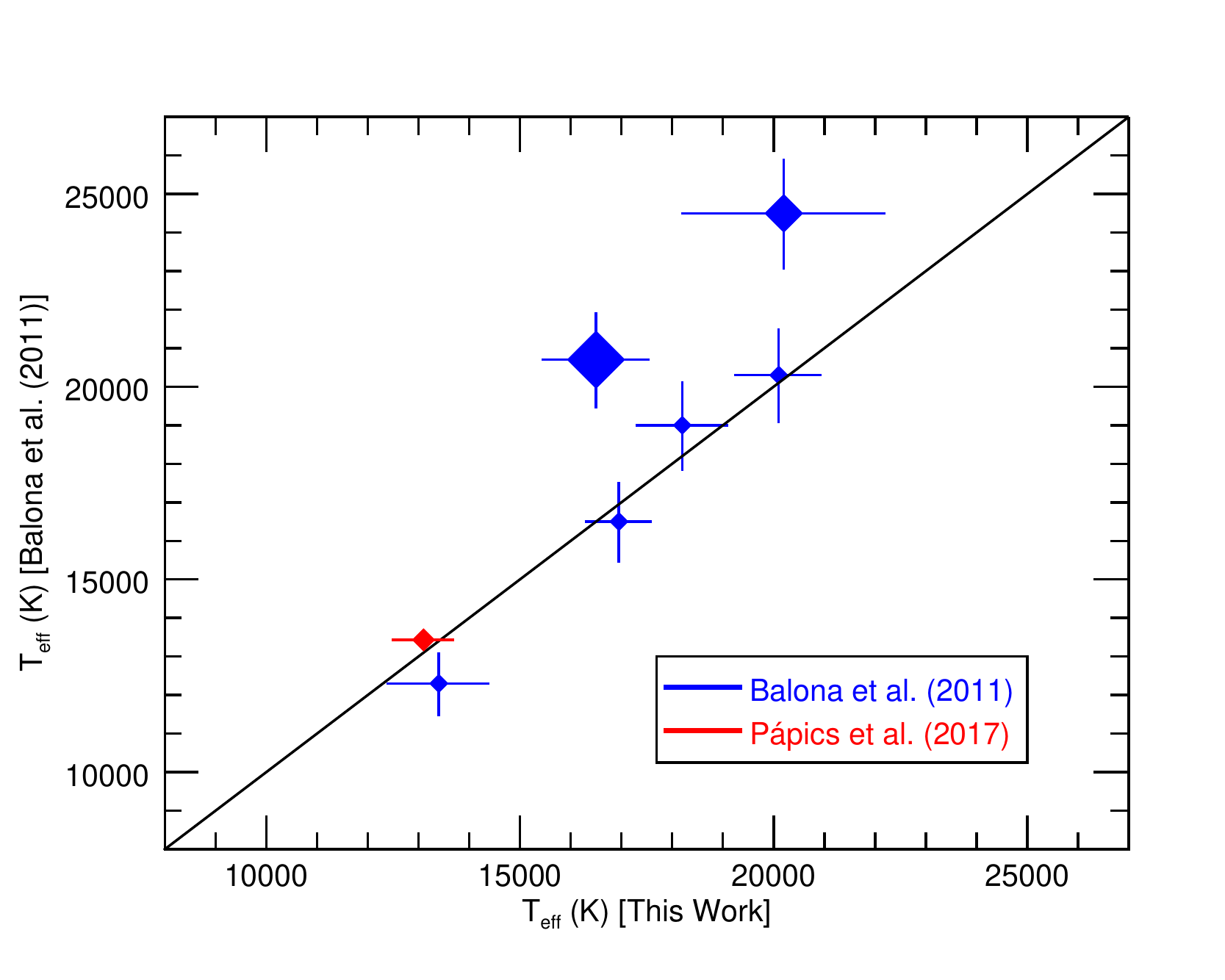}{0.40\textwidth}{(a)}
          \fig{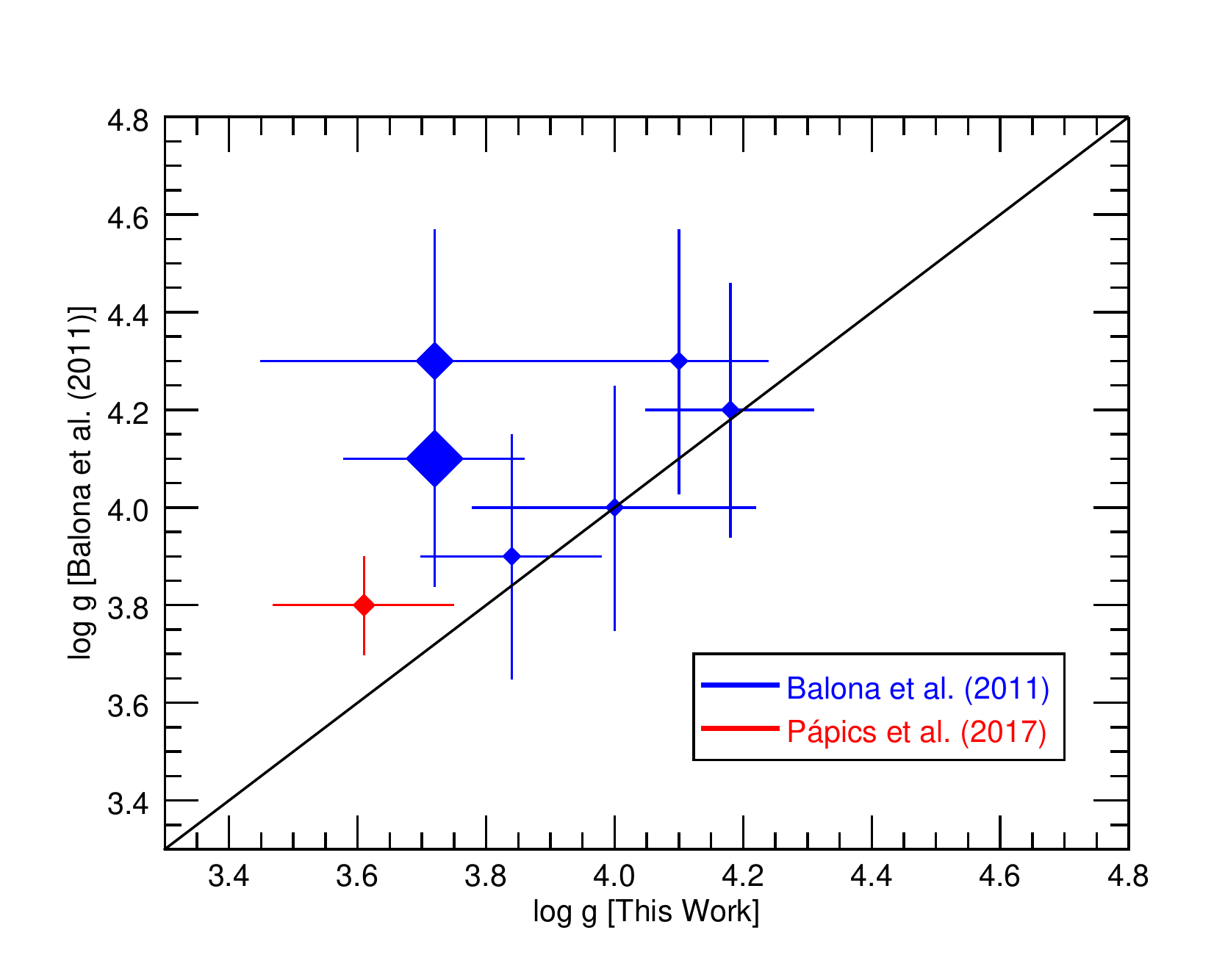}{0.40\textwidth}{(b)}}
\caption{Comparison of our results with \citet{Balona2011} and \citet{Papics2017} for temperature (a) and surface gravity (b). The horizontal and vertical error bars in these figures are the calculated errors from this work, \citet{Balona2011}, and \citet{Papics2017}, respectively. The symbol size is proportional to our measured $V \sin i$ to highlight the discrepancies between our results and those of other works. }
\label{fig:BvH}
\end{figure}

We also include in columns 5 and 6 of Table \ref{tab:Param_comp} some revised measurements from \cite{Balona2011} and \cite{Papics2017}.  \cite{Balona2011} used metal-line blanketed LTE models to model their stars following the methods described by \cite{Ostensen2010}. The biggest reason for the discrepancy with our results and those from \cite{Balona2011} is that they assume $V \sin i  \sim$ 0 km s$^{\rm -1}$ for all of their measurements. As a result, they overestimate $T_{\rm eff}$ and $\log g$ significantly for stars with large $V \sin i$, which leads to wider, shallower hydrogen lines which peak in strength around 10,000 K. \cite{Papics2017} used the BSTAR2006 synthetic spectra \citep{lanz2007} to measure the fundamental parameters of KIC 3459297. They also find $V \sin i$ = 109 $\pm$ 14 km s$^{\rm -1}$ for KIC 3459297, which agrees with our measurements. Figure \ref{fig:BvH} compares our results for $T_{\rm eff}$ and $\log g$ with those from \cite{Balona2011} and \cite{Papics2017}. To emphasize the dependency on $V \sin i$, the sizes of the symbols are proportional to our measured value of $V \sin i$.

Using our measured $T_{\rm eff}$ and $\log g$, we can compare model spectral energy distributions (SED) to photometric data to calculate the radii of our stars using

\begin{equation}
F_\nu / \mathfrak{F}_\nu = ( R_\star / r )^2
\end{equation}

where $F_\nu$ is the apparent monochromatic flux, $\mathfrak{F}_\nu$ is the absolute flux at the surface of the star, and $r$ is the distance to the star. $F_\nu$ was calculated by converting the J, H, and K band magnitudes from the Two Micron All-Sky Survey (2MASS, \cite{Skrutskie2006}) to fluxes using the zero-points from \cite{Cohen2003}. $\mathfrak{F}_\nu$ was determined using model SEDs with our measured $T_{\rm eff}$ and $\log g$. We assumed no interstellar extinction ($A_V$) during this process, as these stars are above the Galactic plane where we would expect low $A_V$ values and they would have negligible effects in the J, H, and K bands. We used the BSTAR2006 models \citep{lanz2007} for stars with $T_{\rm eff} > 15,000$ K and ATLAS models \citep{Castelli2004} for stars with $T_{\rm eff} < 15,000$ K. The distances were calculated by \cite{Jones2018} by converting the parallaxes measured in DR2 using Bayesian statistics. Our error bars were calculated by propagating the errors from $r$, $T_{\rm eff}$ and $\log g$. We then used our measured $\log g$ and $R_\star$ to calculate the masses ($M_\star$) of our stars. KIC 11293898 likely has an underestimated mass, for reasons we discuss later in this work.

\begin{figure}
\plotone{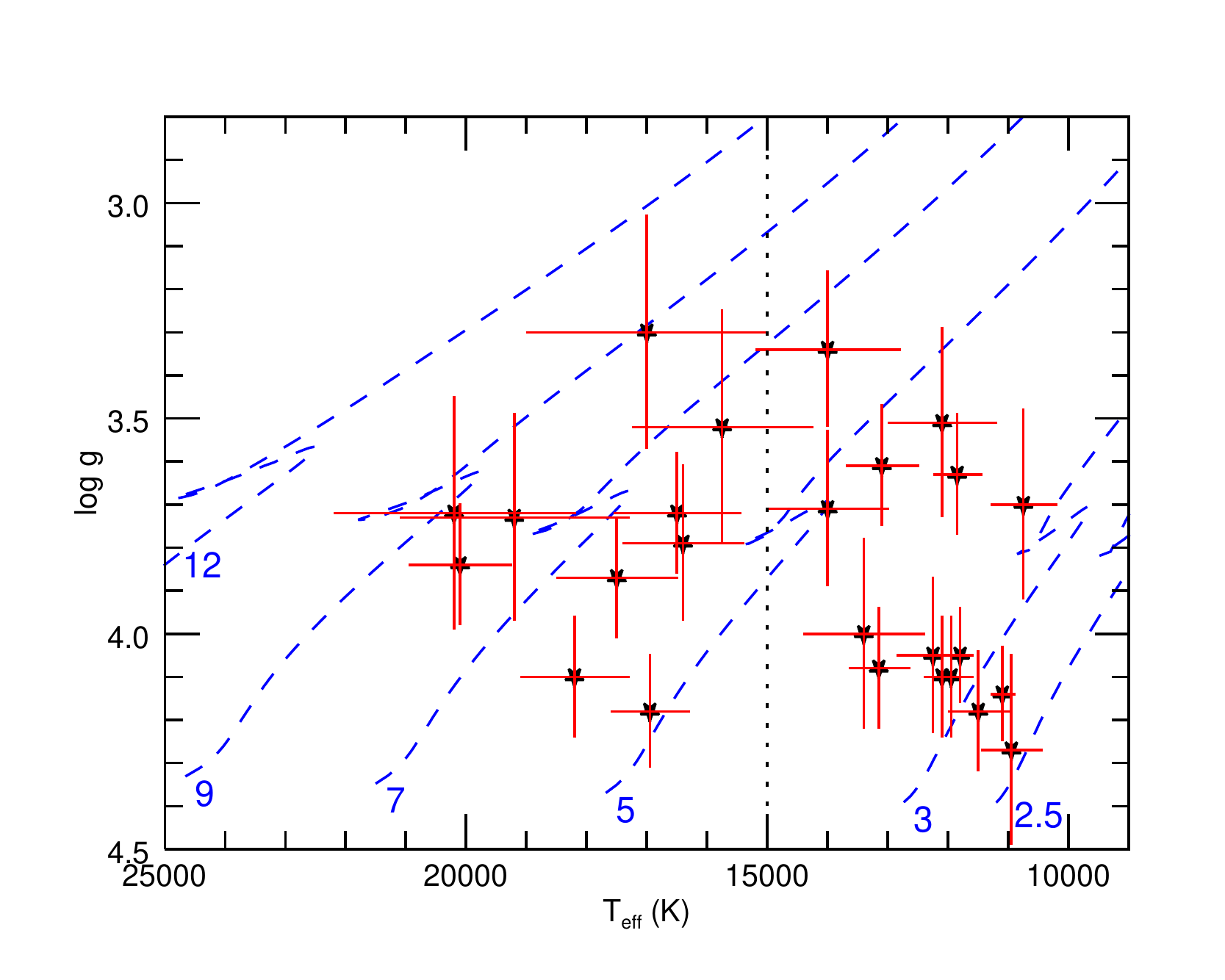}
\caption{HR-Diagram of observed stars using the evolutionary tracks from \cite{ekstrom2012}. The values associated with each track are in solar masses. All stars to the left of the black, dotted line ($T_{\rm eff} > 15,000 K$) were modeled using the BSTAR2006 grid of model spectra. Those on the right ($T_{\rm eff} < 15,000 K$) were modeled using the ATLAS9 models.}
\label{fig:HR}
\end{figure}

Using the non-rotating evolutionary tracks of \cite{ekstrom2012}, we calculated an approximate age of the stars by interpolating between tracks using $T_{\rm eff}$ and $R_\star$. In Table \ref{tab:mm}, we include our measured $R_\star$, $M_\star$ and age ($\tau_\star$), as well as the calculated bolometric luminosity ($L_{\rm bol}$) and the distances from \cite{Jones2018}. We compare our $T_{\rm eff}$ and $\log g$ to the non-rotating evolutionary tracks of \cite{ekstrom2012} in Figure \ref{fig:HR}. Our measured $M_\star$ and $R_\star$ are consistent with their positions along these evolutionary tracks for most stars.

The apparent magnitudes (V), bolometric absolute magnitudes (M$_{bol}$), bolometric corrections (BC), absolute magnitudes ($M_V$), and calculated $A_V$ are all listed in Table \ref{tab:photo}. The $V$ values in column 6 are from the SIMBAD database with an assumed uncertainty of 0.1 mag, and the BC values in column 3 are interpolated from \citet{Flower1996} and \citet{Torres2010}. Using $r$, $V$, and our measurements, we can estimate $A_V$. The extinctions are calculated using 

\begin{equation}
A_V=V - M_V + 5 - 5 \log r
\end{equation}

We find that $A_V$ is consistent with the 3D dust map provided by \cite{Green2018}.

As mentioned before, we noticed our calculated mass for KIC 11293898 to be extremely low compared to its measured effective temperature. Using the non-rotated evolutionary tracks from \cite{ekstrom2012}, we can measure $M_\star$ and $R_\star$ based on $T_{\rm eff}$ and $\log g$. We find $R_\star$ = 5.16 $R_\odot$ and $M_\star$ = 5.98 $M_\odot$ for KIC 11293898. Using this $R_\star$, we find $r$ = 9752 pc, which is well over twice the value calculated by \cite{Jones2018}. 

Our goal with this publication is to improve the measurements of fundamental parameters for pulsating B-type stars in the \textit{Kepler} survey. Using model fitting with the Tlusty BSTAR2006 grid and Kurucz ATLAS9 model atmospheres, as well as use of evolutionary tracks from \cite{ekstrom2012}, we measured $T_{\rm eff}$, $\log g$, $V \sin i$, $M_{\star}$, $\tau_{\star}$, $R_{\star}$, and $L_{\rm bol}$ for 25 pulsating B-type stars . We find that the $T_{\rm eff}$ and $\log g$ measurements from KIC and DR2 are unreliable for these hot stars and that improved stellar parameters are required to continue asteroseismic analysis of these stars.

\acknowledgments
M.\ V.\ McSwain, was supported by NSF grant No.\ AST-1109247.  S.\ W.\ was supported by the National Science Foundation under REU site grant No.\ PHY-1359195.  J.\ L.-B.\ was supported by a Sigma Xi Grant-in-Aid of Research.  A.\ B.\ had support from Kutztown University. This work is also supported by an institutional grant from Lehigh University.  

This work has made use of data from the European Space Agency (ESA) mission {\it Gaia} (\url{https://www.cosmos.esa.int/gaia}), processed by the {\it Gaia} Data Processing and Analysis Consortium (DPAC, \url{https://www.cosmos.esa.int/web/gaia/dpac/consortium}). Funding for the DPAC has been provided by national institutions, in particular the institutions participating in the {\it Gaia} Multilateral Agreement.

This paper includes data collected by the Kepler mission. Funding for the Kepler mission is provided by the NASA Science Mission directorate.

This publication makes use of data products from the Two Micron All Sky Survey, which is a joint project of the University of Massachusetts and the Infrared Processing and Analysis Center/California Institute of Technology, funded by the National Aeronautics and Space Administration and the National Science Foundation.

This research has made use of the SIMBAD database, operated at CDS, Strasbourg, France.

\vspace{5mm}

\facilities{Mayall (RC spectrograph)}

\software{IRAF, IDL}

% Bibliography

\input{Hanes18.bbl}
\input{KICVsini.tex}

\input{KICPhysical_Parameters.v5.tex}

\input{KIC_Param_comp.tex}

\input{KICphoto.v5.tex}

\newpage

\end{document}

%% file: KICVsini.tex
\begin{deluxetable}{ccccccccccc}
\tabletypesize{\scriptsize}
\tablecaption{Projected Rotational Velocities \label{tab:vsini}}
\tablewidth{0pt}
\tablehead{ \colhead{KIC} & \colhead{V $\sin$ i$_{4026}$} & \colhead{$\Delta$V $\sin$ i} & \colhead{V $\sin$ i$_{4388}$} & \colhead{$\Delta$V $\sin$ i} & \colhead{V $\sin$ i$_{4471}$} & \colhead{$\Delta$V $\sin$ i} & \colhead{V $\sin$ i$_{4481}$} & \colhead{$\Delta$V $\sin$ i} & \colhead{V $\sin$ i} & \colhead{$\Delta$V $\sin$ i}\\ \colhead{ID} & \colhead{(km s$^{-1}$)} & \colhead{(km s$^{-1}$)} & \colhead{(km s$^{-1}$)} & \colhead{(km s$^{-1}$)} & \colhead{(km s$^{-1}$)} & \colhead{(km s$^{-1}$)} & \colhead{(km s$^{-1}$)} & \colhead{(km s$^{-1}$)} & \colhead{(km s$^{-1}$)} & \colhead{(km s$^{-1}$)} }
\startdata
1430353  & 190    & 20     & 195    & 10     & 195    & 12     & 235    & 7      & 210 & 26 \\
3459297  & \nodata & \nodata & \nodata & \nodata & 114    & 10     & 130    & 7      & 123 & 12 \\
3839930  & 60     & 5      & 50     & 2      & 50     & 2      & 50     & 2      & 51  & 6  \\
3862353  & 80     & 12     & 55     & 10     & 60     & 5      & 80     & 10     & 64  & 15 \\
4077252  & \nodata & \nodata & \nodata & \nodata & 77     & 22     & 62     & 5      & 65  & 23 \\
4936089  & \nodata & \nodata & \nodata & \nodata & 51     & 6      & 46     & 3      & 48  & 7  \\
4939281  & 115    & 12     & 100    & 5      & 100    & 15     & 135    & 5      & 115 & 20 \\
5477601  & \nodata & \nodata & \nodata & \nodata & 98     & 11     & 86     & 2      & 88  & 11 \\
7630417  & \nodata & \nodata & 135    & 7      & 130    & 7      & 145    & 12     & 135 & 16 \\
8167938  & \nodata & \nodata & \nodata & \nodata & \nodata & \nodata & 70     & 3      & 70  & 3  \\
8264293  & \nodata & \nodata & \nodata & \nodata & 284    & 5      & 283    & 12     & 284 & 13 \\
8381949  & 245    & 10     & 245    & 10     & 245    & 15     & \nodata & \nodata & 245 & 21 \\
8714886  & 75     & 7      & 45     & 5      & 50     & 2      & 50     & 2      & 52  & 9  \\
9227988  & \nodata & \nodata & \nodata & \nodata & 52     & 2      & 48     & 2      & 50  & 3  \\
9278405  & \nodata & \nodata & \nodata & \nodata & 115    & 16     & 108    & 7      & 110 & 17 \\
9468611  & \nodata & \nodata & \nodata & \nodata & 282    & 29     & 258    & 8      & 263 & 30 \\
9715425  & \nodata & \nodata & 125    & 7      & 125    & 7      & 115    & 7      & 122 & 12 \\
9910544  & \nodata & \nodata & \nodata & \nodata & 72     & 4      & 72     & 3      & 72  & 5  \\
9964614  & 85     & 5      & 65     & 5      & 75     & 7      & 85     & 10     & 77  & 14 \\
10118750 & \nodata & \nodata & \nodata & \nodata & \nodata & \nodata & 271    & 16     & 271 & 16 \\
10526294 & \nodata & \nodata & \nodata & \nodata & 37     & 9      & 35     & 4      & 36  & 10 \\
10790075 & \nodata & \nodata & \nodata & \nodata & 74     & 5      & 70     & 2      & 71  & 5  \\
11293898 & 355    & 27     & 360    & 17     & 350    & 15     & \nodata & \nodata & 355 & 35 \\
11360704 & 300    & 7      & 300    & 7      & 300    & 7      & 310    & 7      & 303 & 12 \\
11671923 & \nodata & \nodata & \nodata & \nodata & 89     & 6      & 92     & 5      & 91  & 8 
\enddata
\end{deluxetable}

%% file: KICPhysical_Parameters.v5.tex
\begin{deluxetable}{cccccccccccccc}
\tabletypesize{\scriptsize}
\tablecaption{Physical Parameters \label{tab:mm}}
\tablewidth{0pt}
\tablehead{ \colhead{KIC} & \colhead{T$_{\rm eff}$} & \colhead{$\Delta$T$_{\rm eff}$} &  \colhead{log g} 
& \colhead{$\Delta$log g}  & \colhead{M$_\star$} &  \colhead{R$_\star$}  &  \colhead{L$_{\rm bol}$} & \colhead{$\tau_\star$} & \colhead{r}\\  %%%
\colhead{ID} & \colhead{(K)} & \colhead{(K)} & \colhead{(dex)} & \colhead{(dex)} & \colhead{(M$_\odot$)}  & \colhead{(R$_\odot$)}  & \colhead{(L$_\odot$)} & \colhead{(Myr)} & \colhead{(pc)}}
\startdata
1430353  & 17000 & 2000 & 3.30 & 0.27 & 3.88$^{+2.14}_{-1.73}$ & 7.31$^{+1.76}_{-1.30}$  & 4015$^{+2398}_{-2012}$ & 42$^{+30}_{-16}$   & 7760$^{+1782}_{-1293}$  \\
3459297  & 13100 & 600  & 3.61 & 0.14 & 1.84$^{+1.42}_{-0.56}$ & 3.52$^{+1.33}_{-0.47}$ & 329$^{+253}_{-99}$     & 130$^{+35}_{-22}$  & 3463$^{+599}_{-449}$    \\
3839930  & 16950 & 650  & 4.18 & 0.13 & 4.70$^{+1.22}_{-1.08}$  & 2.92$^{+0.33}_{-0.28}$ & 634$^{+161}_{-142}$    & 32$^{+19}_{-19}$   & 1763$^{+195}_{-161}$    \\
3862353  & 14000 & 1000 & 3.71 & 0.18 & 2.98$^{+1.66}_{-1.23}$ & 3.99$^{+1.05}_{-0.74}$ & 551$^{+313}_{-236}$    & 102$^{+37}_{-28}$  & 9448$^{+2461}_{-1726}$  \\
4077252  & 12100 & 900  & 3.51 & 0.22 & 1.49$^{+0.52}_{-0.50}$  & 3.56$^{+0.48}_{-0.45}$ & 244$^{+86}_{-82}$      & 179$^{+53}_{-51}$  & 2752$^{+236}_{-202}$    \\
4936089  & 12250 & 600  & 4.05 & 0.18 & 1.94$^{+0.42}_{-0.58}$ & 2.18$^{+0.13}_{-0.26}$ & 96$^{+18}_{-27}$       & 103$^{+59}_{-40}$  & 1596$^{+78}_{-71}$      \\
4939281  & 17500 & 1000 & 3.87 & 0.14 & 4.77$^{+1.34}_{-1.17}$ & 4.20$^{+0.51}_{-0.42}$  & 1492$^{+443}_{-393}$   & 49$^{+10}_{-12}$   & 4103$^{+476}_{-389}$    \\
5477601  & 11950 & 350  & 4.10 & 0.14 & 2.39$^{+0.73}_{-0.48}$ & 2.28$^{+0.31}_{-0.16}$ & 96$^{+27}_{-16}$       & 136$^{+43}_{-39}$  & 2134$^{+133}_{-119}$    \\
7630417  & 19200 & 1900 & 3.73 & 0.24 & 4.71$^{+2.73}_{-2.14}$ & 4.91$^{+1.29}_{-0.94}$ & 2947$^{+1779}_{-1428}$ & 33$^{+18}_{-13}$   & 10346$^{+2640}_{-1883}$ \\
8167938  & 13400 & 1000 & 4.00 & 0.22 & 9.14$^{+2.64}_{-3.75}$ & 5.01$^{+0.46}_{-0.86}$ & 728$^{+211}_{-299}$    & 111$^{+51}_{-25}$  & 1859$^{+153}_{-133}$    \\
8264293  & 13150 & 500  & 4.08 & 0.14 & 2.62$^{+1.61}_{-0.48}$ & 2.44$^{+0.73}_{-0.14}$ & 161$^{+98}_{-26}$      & 105$^{+34}_{-34}$  & 1328$^{+70}_{-64}$      \\
8381949  & 20200 & 2000 & 3.72 & 0.27 & 6.74$^{+3.46}_{-2.84}$ & 5.94$^{+1.30}_{-0.96}$  & 5282$^{+2796}_{-2320}$ & 28$^{+12}_{-9}$    & 4019$^{+840}_{-603}$    \\
8714886  & 18200 & 900  & 4.10 & 0.14 & 3.47$^{+0.72}_{-0.67}$ & 2.75$^{+0.21}_{-0.18}$ & 747$^{+159}_{-148}$    & 16$^{+15}_{-12}$   & 1560$^{+108}_{-95}$     \\
9227988  & 14000 & 1200 & 3.34 & 0.18 & 2.57$^{+1.37}_{-1.07}$ & 5.68$^{+1.42}_{-1.06}$ & 1114$^{+627}_{-505}$   & 93$^{+50}_{-31}$   & 6799$^{+1642}_{-1174}$  \\
9278405  & 11100 & 200  & 4.14 & 0.11 & 2.11$^{+0.31}_{-0.31}$ & 2.05$^{+0.10}_{-0.10}$   & 57$^{+6}_{-6}$         & 173$^{+35}_{-35}$  & 654$^{+19}_{-18}$       \\
9468611  & 11200 & 600  & 3.90 & 0.22 & 1.85$^{+0.80}_{-0.55}$  & 2.53$^{+0.47}_{-0.25}$ & 90$^{+37}_{-23}$       & 224$^{+66}_{-76}$  & 2729$^{+310}_{-254}$    \\
9715425  & 15750 & 750  & 3.52 & 0.27 & 2.14$^{+1.24}_{-0.97}$ & 4.21$^{+1.08}_{-0.77}$ & 982$^{+523}_{-386}$    & 67$^{+30}_{-22}$   & 6059$^{+1547}_{-1101}$  \\
9910544  & 12100 & 300  & 4.10 & 0.14 & 4.21$^{+0.80}_{-0.76}$  & 3.03$^{+0.19}_{-0.17}$ & 177$^{+26}_{-24}$      & 184$^{+16}_{-18}$  & 1294$^{+79}_{-70}$      \\
9964614  & 20100 & 850  & 3.84 & 0.14 & 6.62$^{+2.39}_{-1.95}$ & 5.13$^{+0.85}_{-0.66}$ & 3862$^{+1371}_{-1109}$ & 29$^{+3}_{-5}$     & 3407$^{+561}_{-427}$    \\
10118750 & 10950 & 500  & 4.27 & 0.22 & 2.94$^{+0.99}_{-0.89}$ & 2.08$^{+0.26}_{-0.21}$ & 56$^{+16}_{-14}$       & 196$^{+98}_{-110}$ & 3319$^{+405}_{-329}$    \\
10526294 & 11500 & 500  & 4.18 & 0.14 & 3.26$^{+0.81}_{-0.73}$ & 2.43$^{+0.25}_{-0.21}$ & 93$^{+23}_{-20}$       & 189$^{+58}_{-66}$  & 3025$^{+297}_{-251}$    \\
10790075 & 11850 & 400  & 3.63 & 0.14 & 2.28$^{+0.92}_{-0.77}$ & 3.83$^{+0.72}_{-0.59}$ & 260$^{+101}_{-84}$     & 184$^{+34}_{-43}$  & 4153$^{+699}_{-536}$    \\
11293898 & 16400 & 1000 & 3.79 & 0.18 & 0.92$^{+0.28}_{-0.25}$ & 2.02$^{+0.25}_{-0.21}$ & 266$^{+82}_{-74}$      & 58$^{+18}_{-10}$   & 3823$^{+459}_{-374}$    \\
11360704 & 16500 & 1050 & 3.72 & 0.14 & 7.26$^{+2.55}_{-2.08}$ & 6.16$^{+0.99}_{-0.77}$ & 2532$^{+947}_{-797}$   & 52$^{+22}_{-12}$   & 3428$^{+537}_{-414}$    \\
11671923 & 11800 & 200  & 4.05 & 0.11 & 2.96$^{+0.50}_{-0.50}$   & 2.69$^{+0.17}_{-0.17}$ & 127$^{+17}_{-17}$      & 188$^{+17}_{-19}$  & 1065$^{+53}_{-49}$
\enddata   
\end{deluxetable}

%% file: KIC_Param_comp.tex
\begin{deluxetable}{@{\extracolsep{4pt}}rrrrrrrr}
\tabletypesize{\scriptsize}
\tablecaption{Comparison of Stellar Parameters of \textit{Kepler} B Stars \label{tab:Param_comp}}
\tablewidth{0pt}
\tablehead{
\colhead{} & \multicolumn{2}{c}{KIC} & \multicolumn{1}{c}{DR2} & \multicolumn{2}{c}{Other} & \multicolumn{2}{c}{This Work}\\
\cline{2-3}\cline{4-4}\cline{5-6}\cline{7-8}
\colhead{KIC} & \colhead{$T_{\rm eff}$} &  \colhead{$\log g$} & \colhead{$T_{\rm eff}$} & \colhead{$T_{\rm eff}$} &  \colhead{$\log g$} & \colhead{$T_{\rm eff}$} &  \colhead{$\log g$} \\
\colhead{ID} & \colhead{(K)} & \colhead{(dex)} & \colhead{(K)} & \colhead{(K)} & \colhead{(dex)} & \colhead{(K)} & \colhead{(dex)}
}
\startdata
1430353  & 10765 & 3.653 & 9585$^{+106}_{-73}$    &\nodata         &\nodata        & 17000 $\pm$ 2000 & 3.30 $\pm$ 0.27  \\
3459297  & 10592 & 4.629 & 9365$^{+248}_{-3148}$  & 13430 $\pm$ 250\tablenotemark{a}  & 3.8 $\pm$ 0.1\tablenotemark{a}   & 13100 $\pm$ 600  & 3.61 $\pm$ 0.14 \\
3839930  & 11272 & 4.277 & 8712$^{+774}_{-1039}$  & 16500 $\pm$ 1000\tablenotemark{b} & 4.2 $\pm$ 0.3\tablenotemark{b}  & 16950 $\pm$ 650  & 4.18 $\pm$ 0.13 \\
3862353  & 10738 & 4.025 & 9623$^{+194}_{-733}$   &\nodata         &\nodata        & 14000 $\pm$ 1000 & 3.71 $\pm$ 0.18 \\
4077252  & 10514 & 3.948 & 9312$^{+282}_{-283}$   &\nodata         &\nodata        & 12100 $\pm$ 900  & 3.51 $\pm$ 0.22 \\
4936089  & 11295 & 4.275 & 9554$^{+101}_{-2140}$  &\nodata         &\nodata        & 12250 $\pm$ 600  & 4.05 $\pm$ 0.18 \\
4939281  & 11298 & 4.316 & 8726$^{+943}_{-3928}$  &\nodata         &\nodata        & 17500 $\pm$ 1000 & 3.87 $\pm$ 0.14 \\
5477601  & 10906 & 4.138 & 9304$^{+275}_{-227}$   &\nodata         &\nodata        & 11950 $\pm$ 350  & 4.10 $\pm$ 0.14  \\
7630417  & 10449 & 4.581 & 9613$^{+128}_{-90}$    &\nodata         &\nodata        & 19200 $\pm$ 1900 & 3.73 $\pm$ 0.24 \\
8167938  & 11167 & 4.228 & 9453$^{+315}_{-3945}$  & 12300 $\pm$ 800\tablenotemark{b}  & 4.0 $\pm$ 0.3\tablenotemark{b}    & 13400 $\pm$ 1000 & 4.00 $\pm$ 0.22    \\
8264293  & 10038 & 4.458 & 9530$^{+170}_{-2116}$  &\nodata         &\nodata        & 13150 $\pm$ 500  & 4.08 $\pm$ 0.14 \\
8381949  & 9782  & 4.394 & 9020$^{+680}_{-4222}$  & 24500 $\pm$ 1400\tablenotemark{b} & 4.3 $\pm$ 0.3\tablenotemark{b} & 20200 $\pm$ 2000 & 3.72 $\pm$ 0.27 \\
8714886  & 9142  & 4.093 & 9351$^{+260}_{-209}$   & 19000 $\pm$ 1200\tablenotemark{b} & 4.3 $\pm$ 0.3\tablenotemark{b} & 18200 $\pm$ 900  & 4.10 $\pm$ 0.14  \\
9227988  & 10890 & 4.023 & 9546$^{+60}_{-2301}$   &\nodata         &\nodata        & 14000 $\pm$ 1200 & 3.34 $\pm$ 0.18 \\
9278405  & 11486 & 4.199 & 8614$^{+1048}_{-1687}$ &\nodata         &\nodata        & 11100 $\pm$ 200  & 4.14 $\pm$ 0.11 \\
9468611  & 11063 & 3.735 & 9365$^{+248}_{-3148}$  &\nodata         &\nodata        & 10750 $\pm$ 550  & 3.70 $\pm$ 0.22  \\
9715425  & 11199 & 4.114 & 8728$^{+964}_{-3930}$  &\nodata         &\nodata        & 15750 $\pm$ 1500 & 3.52 $\pm$ 0.27 \\
9910544  & 10698 & 4.079 & 9540$^{+202}_{-629}$   &\nodata         &\nodata        & 12100 $\pm$ 300  & 4.10 $\pm$ 0.14  \\
9964614  & 8915  & 4.067 & 8869$^{+724}_{-906}$   & 20300 $\pm$ 1200\tablenotemark{b} & 3.9 $\pm$ 0.2\tablenotemark{b} & 20100 $\pm$ 850  & 3.84 $\pm$ 0.14 \\
10118750 & 11147 & 3.751 & 9045$^{+521}_{-2663}$  &\nodata         &\nodata        & 10950 $\pm$ 500  & 4.27 $\pm$ 0.22 \\
10526294 & 11072 & 3.743 & 8983$^{+743}_{-1563}$  &\nodata         &\nodata        & 11500 $\pm$ 500  & 4.18 $\pm$ 0.14 \\
10790075 & 11396 & 3.771 & 9055$^{+637}_{-2005}$  &\nodata         &\nodata        & 11850 $\pm$ 400  & 3.63 $\pm$ 0.14 \\
11293898 & 15072 & 4.932 & 8599$^{+994}_{-1179}$  &\nodata         &\nodata        & 16400 $\pm$ 1000 & 3.79 $\pm$ 0.18 \\
11360704 & 12400 & 4.934 & 8869$^{+724}_{-906}$   & 20700 $\pm$ 1200\tablenotemark{b} & 4.1 $\pm$ 0.3\tablenotemark{b} & 16500 $\pm$ 1050 &3.72 $\pm$ 0.14 \\
11671923 & 10044 & 4.237 & 9285$^{+441}_{-1643}$  &\nodata         &\nodata        & 11800 $\pm$ 200  & 4.05 $\pm$ 0.11
\enddata
\tablenotetext{a}{\citealt{Papics2017}}
\tablenotetext{b}{\citealt{Balona2011}}
\end{deluxetable}

%% file: KICphoto.v5.tex
\begin{deluxetable}{cccccccccc}
\tabletypesize{\scriptsize}
\tablecaption{Photometry \label{tab:photo}}
\tablewidth{0pt}
\tablehead{ \colhead{KIC} & \colhead{M$_{\rm bol}$} & \colhead{BC} & \colhead{$\Delta$BC} & \colhead{M$_V$} & \colhead{V} & \colhead{A$_V$}  \\ \colhead{ID} &  \colhead{(mag)} & \colhead{(mag)} & \colhead{(mag)} & \colhead{(mag)} & \colhead{(mag)} & \colhead{(mag)} }
\startdata
1430353  & -4.28$^{+1.45}_{-1.21}$ & -1.53 & 0.61 & -2.75$^{+1.57}_{-1.36}$ & 12.71  & 1.01$^{+1.95}_{-1.6}$  \\
3459297  & -1.56$^{+0.93}_{-0.73}$ & -0.90 & 0.27 & -0.66$^{+0.97}_{-0.78}$ & \nodata & \nodata                        \\
3839930  & -2.27$^{+0.62}_{-0.54}$ & -1.52 & 0.20 & -0.75$^{+0.65}_{-0.58}$ & 10.79  & 0.31$^{+0.86}_{-0.73}$     \\
3862353  & -2.12$^{+1.41}_{-1.06}$ & -1.07 & 0.41 & -1.05$^{+1.47}_{-1.14}$ & \nodata & \nodata                     \\
4077252  & -1.24$^{+0.7}_{-0.67}$  & -0.70 & 0.43 & -0.54$^{+0.83}_{-0.8}$  & 12.29  & 0.63$^{+0.93}_{-0.88}$   \\
4936089  & -0.23$^{+0.44}_{-0.43}$ & -0.73 & 0.29 & 0.51$^{+0.53}_{-0.52}$  & 11.93  & 0.41$^{+0.58}_{-0.56}$   \\
4939281  & -3.2$^{+0.72}_{-0.64}$  & -1.60 & 0.30 & -1.61$^{+0.78}_{-0.7}$  & \nodata & \nodata                  \\
5477601  & -0.22$^{+0.38}_{-0.35}$ & -0.67 & 0.17 & 0.45$^{+0.42}_{-0.39}$  & \nodata & \nodata                          \\
7630417  & -3.94$^{+1.48}_{-1.17}$ & -1.80 & 0.50 & -2.14$^{+1.56}_{-1.28}$ & \nodata & \nodata                  \\
8167938  & -2.43$^{+0.69}_{-0.66}$ & -0.96 & 0.43 & -1.46$^{+0.82}_{-0.79}$ & 10.77  & 0.89$^{+0.92}_{-0.87}$    \\
8264293  & -0.79$^{+0.39}_{-0.37}$ & -0.91 & 0.22 & 0.13$^{+0.45}_{-0.43}$  & 11.26  & 0.52$^{+0.52}_{-0.5}$    \\
8381949  & -4.58$^{+1.28}_{-1.06}$ & -1.91 & 0.50 & -2.66$^{+1.38}_{-1.17}$ & 10.96  & 0.6$^{+1.73}_{-1.39}$   \\
8714886  & -2.45$^{+0.51}_{-0.48}$ & -1.68 & 0.25 & -0.77$^{+0.57}_{-0.54}$ & 10.86  & 0.66$^{+0.66}_{-0.62}$  \\
9227988  & -2.89$^{+1.37}_{-1.08}$ & -1.07 & 0.49 & -1.82$^{+1.45}_{-1.18}$ & 12.54  & 0.19$^{+1.89}_{-1.46}$ \\
9278405  & 0.33$^{+0.2}_{-0.19}$   & -0.49 & 0.10 & 0.82$^{+0.22}_{-0.22}$  & 10.16  & 0.26$^{+0.27}_{-0.26}$      \\
9468611  & -0.16$^{+0.7}_{-0.61}$  & -0.51 & 0.30 & 0.35$^{+0.76}_{-0.69}$  & \nodata & \nodata                                 \\
9715425  & -2.75$^{+1.46}_{-1.16}$ & -1.35 & 0.51 & -1.4$^{+1.55}_{-1.26}$  & \nodata & \nodata                                \\
9910544  & -0.89$^{+0.36}_{-0.33}$ & -0.70 & 0.14 & -0.19$^{+0.39}_{-0.36}$ & 10.80  & 0.43$^{+0.49}_{-0.45}$   \\
9964614  & -4.24$^{+0.88}_{-0.7}$  & -1.90 & 0.22 & -2.33$^{+0.91}_{-0.73}$ & 10.61  & 0.28$^{+1.23}_{-0.97}$    \\
10118750 & 0.36$^{+0.7}_{-0.6}$    & -0.46 & 0.25 & 0.81$^{+0.74}_{-0.65}$  & \nodata & \nodata                           \\
10526294 & -0.19$^{+0.59}_{-0.53}$ & -0.58 & 0.25 & 0.38$^{+0.64}_{-0.58}$  & \nodata & \nodata                         \\
10790075 & -1.31$^{+0.88}_{-0.69}$ & -0.65 & 0.20 & -0.66$^{+0.9}_{-0.72}$  & 12.96  & 0.53$^{+1.23}_{-0.97}$     \\
11293898 & -1.33$^{+0.76}_{-0.67}$ & -1.45 & 0.32 & 0.12$^{+0.82}_{-0.74}$  & \nodata & \nodata                       \\
11360704 & -3.78$^{+0.92}_{-0.77}$ & -1.46 & 0.34 & -2.32$^{+0.98}_{-0.84}$ & 10.62  & 0.26$^{+1.25}_{-1.03}$     \\
11671923 & -0.53$^{+0.28}_{-0.26}$ & -0.64 & 0.10 & 0.11$^{+0.3}_{-0.28}$   & 10.58  & 0.33$^{+0.39}_{-0.36}$   
\enddata
\end{deluxetable}